\documentclass[]{spie}  %>>> use for US letter paper
\usepackage[]{graphicx}
\usepackage{url}

\title{AXTAR: Mission Design Concept} 

\author{Paul S.\ Ray\supit{a}, Deepto Chakrabarty\supit{b}, Colleen
  A.\,Wilson-Hodge\supit{c},  Bernard F.\, Phlips\supit{a}, \\
  Ronald A.\,Remillard\supit{b}, Alan M.\,Levine\supit{b}, 
  Kent S.\,Wood\supit{a}, Michael T.\ Wolff\supit{a}, Chul S.\,Gwon\supit{a},
  Tod E.\,Strohmayer\supit{d}, Michael Baysinger\supit{e}, Michael S.\,Briggs\supit{c},  
  Peter
  Capizzo\supit{e}, Leo Fabisinski\supit{e},
  Randall
  C.\, Hopkins\supit{e}, Linda S. Hornsby\supit{e}, Les
  Johnson\supit{e}, C. Dauphne Maples\supit{e}, \\ Janie H.\,Miernik\supit{e}, Dan Thomas\supit{e}, Gianluigi De
  Geronimo\supit{f} 
\skiplinehalf
\supit{a}Space Science Division, Naval Research Laboratory,
Washington, DC 20375, USA\\ 
\supit{b}Kavli Institute for Astrophysics and Space Research, Massachusetts Institute of Technology, Cambridge, MA 02139, USA\\
\supit{c}Space Science Office, NASA Marshall Space Flight Center,
Huntsville, AL 35812, USA\\ 
\supit{d}NASA Goddard Space Flight Center, Greenbelt, MD 20771, USA \\
\supit{e}Advanced Concepts Office, NASA Marshall Space Flight Center,
Huntsville, AL 35812, USA \\ 
\supit{f}Instrumentation Division, Brookhaven National Laboratory, Upton, NY 11973, USA\\
}

\authorinfo{Send correspondence to P.S.R. E-mail: paul.ray@nrl.navy.mil}
\begin{document} 
  \maketitle 

%%%%%%%%%%%%%%%%%%%%%%%%%%%%%%%%%%%%%%%%%%%%%%%%%%%%%%%%%%%%% 
\begin{abstract}
The Advanced X-ray Timing Array (AXTAR) is a mission
concept for X-ray timing of compact objects that combines very large
collecting area, broadband spectral coverage, high time resolution,
highly flexible scheduling, and an ability to respond promptly to
time-critical targets of opportunity. It is optimized for
submillisecond timing of bright Galactic X-ray sources in order to
study phenomena at the natural time scales of neutron star surfaces
and black hole event horizons, thus probing the physics of ultradense
matter, strongly curved spacetimes, and intense magnetic
fields. AXTAR's main instrument, the Large Area Timing Array (LATA) is a collimated instrument with 2--50 keV coverage and over 3 square meters effective area. The LATA is made up of an array of supermodules that house 2-mm thick silicon pixel detectors. AXTAR will provide a significant improvement in
effective area (a factor of 7 at 4 keV and a factor of 36 at 30 keV) over the RXTE PCA. AXTAR will also carry a sensitive Sky
Monitor (SM) that acts as a trigger for pointed observations of X-ray
transients in addition to providing high duty cycle monitoring of the
X-ray sky. We review the science goals and technical concept for AXTAR
and present results from a preliminary mission design study.
\end{abstract}

%>>>> Include a list of keywords after the abstract 

\keywords{Neutron Stars, Black Holes, X-ray Timing, Silicon Pixel Detectors, Mission Concepts}

%%%%%%%%%%%%%%%%%%%%%%%%%%%%%%%%%%%%%%%%%%%%%%%%%%%%%%%%%%%%%
\section{INTRODUCTION}
\label{sec:intro}  % \label{} allows reference to this section

The properties of ultradense matter and strongly curved spacetime and
the behavior of matter and radiation in the extreme environments near
compact objects are among the most fundamental problems
in astrophysics.  X-ray timing measurements have powerful advantages
for studying these problems \cite{lam04}.  The X-ray band contains
most of the power emitted by accreting neutron stars and black holes,
and this radiation is relatively penetrating even in these complex
environments.  The rapid X-ray variability of these objects encodes
their basic physical parameters, and interpretation of this
variability is relatively straightforward for rotating or orbital
origins.  In many cases, the properties of the X-ray variability
allow extremely precise measurements and detailed quantitative
inferences. 
The scientific promise of X-ray timing has been spectacularly demonstrated
by the success of NASA's Rossi X-ray Timing Explorer (RXTE; effective area
$A_{\rm eff}=0.6$ m$^2$; launched 1995), which has revealed an extraordinary
range of previously unknown variability phenomena from neutron stars and 
black holes.  However, redeeming that promise and exploiting these phenomena
to answer fundamental astrophysical questions will require a larger-area
follow-on mission.  A detailed scientific case for such a mission was first
explored at the conference {\em X-ray Timing 2003: Rossi and Beyond} in
Cambridge, Massachusetts\cite{rossi03}. 

\begin{figure}
\begin{center}
\includegraphics[width=5.25in]{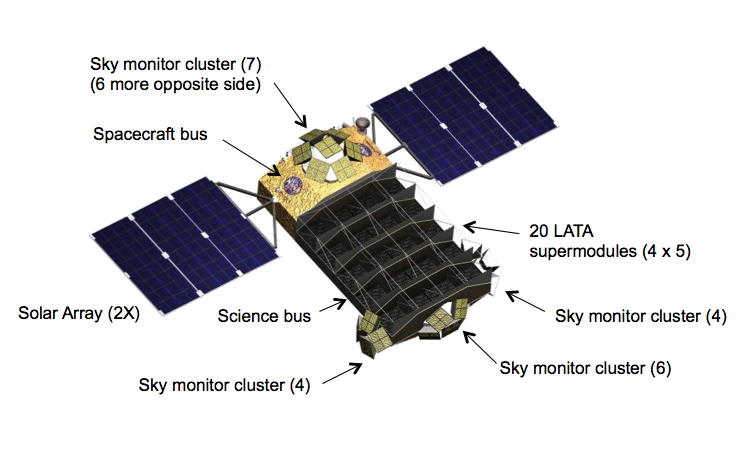}
\end{center}
\caption{AXTAR spacecraft configuration with 20 LATA
  supermodules. This configuration is within the volume and payload
  mass limits for a Taurus~II launcher, and will also easily work with
  a Falcon 9 launcher. \label{fig:TaurusII_config}}
\end{figure}

\begin{figure}[b]
\begin{center}
\includegraphics[width=3.5in]{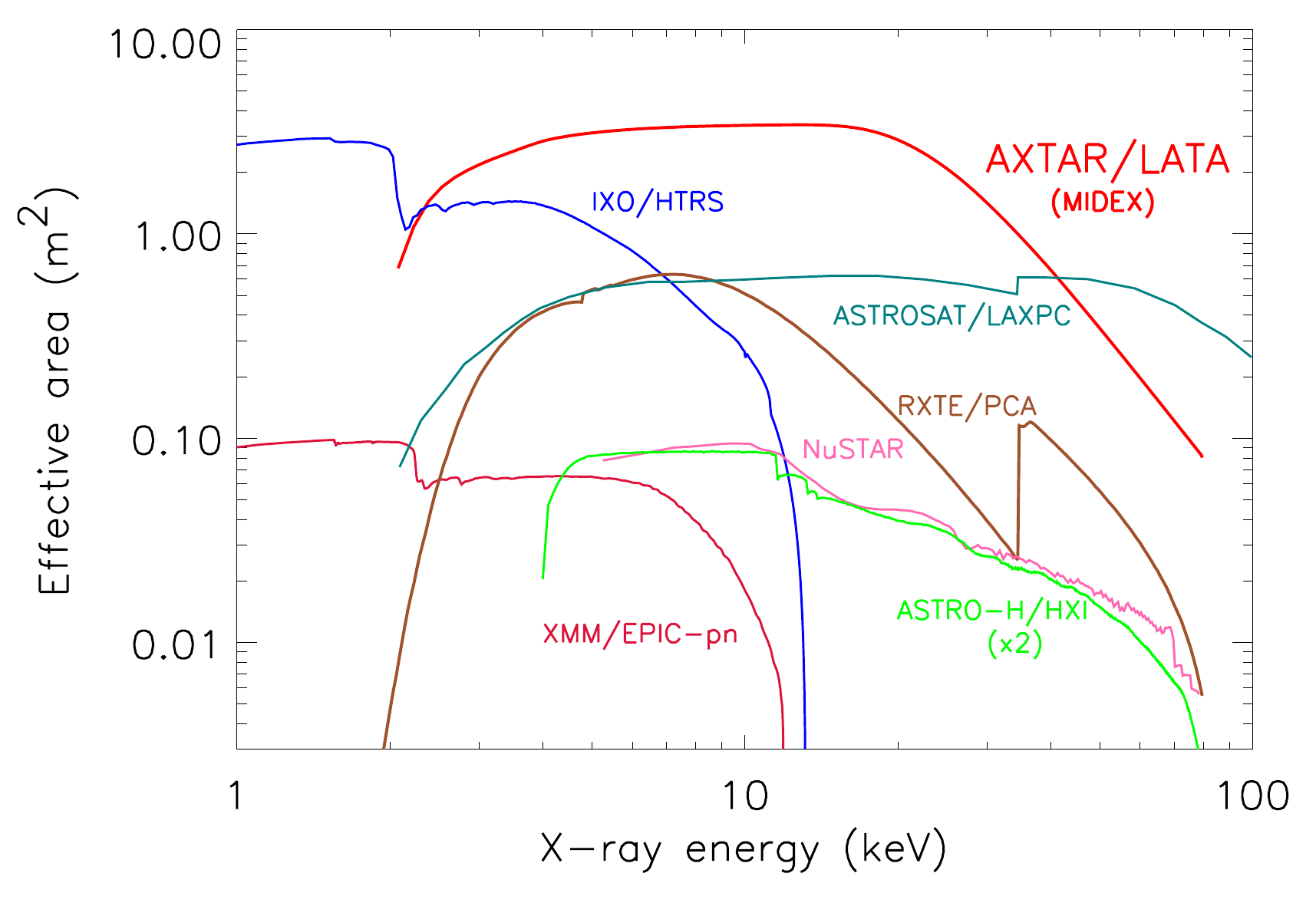}
\end{center}
\caption{Effective area as a function of energy of the AXTAR LATA baseline detector concept. Several other current and planned missions are shown for comparison.\label{fig:effarea}}
\end{figure}

In this paper we describe the Advanced X-ray Timing Array (AXTAR; see 
Figure~\ref{fig:TaurusII_config}), a new mission concept with significantly larger effective area than
RXTE (see Figure \ref{fig:effarea}), allowing it to exploit the phenomena discovered by RXTE to measure
fundamental parameters of neutron stars and black holes and to study a
broad range of topics in high-energy astrophysics.  AXTAR was
originally proposed as a medium-class probe concept in the 2007 NASA
Astrophysics Strategic Mission Concept Study call.  More recently, we
have been developing AXTAR as a NASA Explorer-class mission concept.
AXTAR's baseline main instrument, the Large Area Timing Array (LATA), is a collimated instrument employing thick Si pixel
detectors that achieves 2--50 keV energy coverage and over 3~m$^2$ effective area.  The
mission also carries a sensitive Sky Monitor (SM) that acts as a trigger
for pointed observations of X-ray transients in addition to providing
high--duty-cycle monitoring of the X-ray sky.
Several other large-area follow-on mission concepts have also been previously
proposed\cite{kaa04,elv06,bvs+01,bar08}.

We begin with a brief overview of the science objectives and mission drivers
for AXTAR, then discuss our detector and mission concepts, and finally 
summarize the results of our recent mission design study.

\section{SCIENCE GOALS AND MISSION REQUIREMENTS}
\label{sec:sci}  % \label{} allows reference to this section

The natural time scales near stellar mass black holes (BHs) and
neutron stars (NSs) (either the free-fall time or the fastest stable
orbit time) are in the millisecond range.  These time scales
characterize the fundamental physical properties of compact objects:
mass, radius, and angular momentum.  For example, the maximum spin
rate of a neutron star is set by the equation of state of the ultradense matter
in its interior, a fundamental property of matter that still eludes
us.  Similarly, orbital periods at a given radius near a black hole are set by
the black hole's mass, angular momentum, and the laws of relativistic gravity.
Since its 1995 launch, RXTE has discovered millisecond oscillations
from neutron stars that trace their spin rate and millisecond oscillations from
accreting black holes with frequencies that scale inversely with black hole mass and
are consistent with the orbital time scale of matter moving in the
strongly curved spacetime near the black hole event horizon.  However, while
RXTE revealed the existence of these phenomena, it lacks the
sensitivity to fully exploit them in determining the fundamental
properties of neutron stars and black holes (see Ref.~\citenum{rossi03} and papers
therein). 

The large sensitivity gain of AXTAR's Large Area Timing Array will
permit the exploration of multiple topics in the physics and astrophysics of
compact objects, including:
\begin{itemize}
\item{{\bf Neutron star mass, radius, and equation of state.}  AXTAR will
  extract information encoded in the pulse shapes of thermonuclear
  X-ray burst oscillations\cite{nss02,bsm+05}, kilohertz
  quasi-periodic oscillations\cite{mlp98,sv99} (QPOs), and accreting
  millisecond X-ray pulsars\cite{pg03,pc99} to measure or constrain
  the fundamental propeties of neutron stars.  Asteroseismology 
  of neutron star internal structure using X-ray oscillations during
  magnetar giant bursts will also be possible\cite{sw06,ws06}.}
\item{{\bf Black holes and the physics of strongly curved spacetime.}
  AXTAR will explore the physical origin of the stable, high-frequency
  QPOs observed in accreting black holes\cite{rm06}, testing their
  ties to black hole mass and spin.  With its large effective area,
  broad energy coverage, and good energy resolution, AXTAR will also
  be an ideal instrument for time-resolved spectrophotometric study of
  relativistically broadened iron fluorescence lines in both black
  holes\cite{rn03} and neutron stars\cite{cmh+09}, as well as
  phase-resolved spectroscopy of low-frequency QPOs.}  
\item{{\bf Physics of nuclear burning and exotic nucleosynthesis.}
 AXTAR will study thermonuclear X-ray bursts and superbursts from
 neutron stars with unprecedented sensitivity, allowing detailed
 time-resolution of burst rises, hot spot evolution, and the
 evolution of both early and late-time burst oscillations\cite{sb06}.
 AXTAR will be sensitive to spectral absorption edges from exotic heavy
 metals produced in the nuclear burning chains of radius-expansion
 bursts\cite{wbs06}.  AXTAR will also continue exploration of the mHz QPOs 
 (interpreted as unstable nuclear burning on the neutron star surface) 
 well beyond the limited instances accessible to RXTE\cite{avw+08}.}
%\item{{\bf Accretion disk and jet physics.}}
%\item{{\bf Magnetars and extreme magnetic fields.}}
\item{{\bf Pushing the envelope.} AXTAR will provide clear detection
 or refutation of effects that were hinted at in RXTE data but remain
 unconfirmed. A leading example is provided by photon bubble 
 oscillations\cite{kaj+96}, a radiation hydrodynamic effect predicted in 
 accreting pulsars, but thus far detected with only marginal and disputed 
 significance\cite{jka00}. If detected, these oscillations will probe the interaction of high intensity radiation with matter in accretion columns where the mass accreted per unit area far exceeds the Eddington limit. }
\end{itemize}

The AXTAR Sky Monitor will monitor hundreds of X-ray sources in
addition to serving as a trigger for target-of-opportunity
observations of active X-ray transients.  The activity level of
Galactic X-ray sources is inherently variable on time scales from
milliseconds to days to years.  While RXTE demonstrated the need
for a sky monitor as a trigger for pointed observations and to provide
spectral and activity context about sources between pointed
observations, we note that a sensitive sky monitor will serve as a
primary science instrument for a wide variety of investigations,
including:
\begin{itemize}
\item{Daily monitoring of the flux and spin history of $\sim 50$
  accretion-powered X-ray pulsars over 5 years, hence testing the
  theory of magnetic disk accretion and angular momentum
  transfer\cite{bcc+97}.  Dynamical measurement of orbital parameters
  through pulse timing for all of these systems\cite{bcc+97}.}
\item{Phase-coherent tracking of the spin down of magnetars\cite{gk02} using multi-day integrations and
  detection of rotational glitches\cite{klc00,kgw+03}, providing a
  window into neutron star interiors in the presence of superstrong magnetic
  fields. Detection of giant flares from magnetars in the Local
  Group\cite{sw06,ws06}.} 
\item{Long-term variability monitoring of the brightest active
  galactic nuclei (AGN) over hours, days, and years, testing how the
  variability timescale relates to black hole mass and this relationship
  extends to stellar-mass black holes\cite{mkk+06}.}
\item{Identifying thermonuclear X-ray bursters entering a frequent
  bursting state.  Detecting rare thermonuclear ``superbursts'' (deep
  nuclear burning events) as they occur, anywhere in the
  Galaxy. Statistical studies of thermonuclear burst activity over a
  wide range of mass accretion rates\cite{sb06,gmh+08}.}
\item{Long-term timing of superorbital periodicities and accretion
  disk precession in low-mass X-ray binaries\cite{wlc+06}
  (LMXBs). Orbital period evolution of eclipsing
  systems\cite{whw+02}.}
\item{Detection of gamma-ray bursts, X-ray flashes, and other
  short-timescale, isolated, or very low duty cycle flares.}
\end{itemize}

In addition, the sky monitor offers partnership with observatories for
wide-angle astronomy in other wavebands. New facilities are under
construction for wide-angle radio monitoring such as LOFAR (30--240
MHz), the MWA (80--240 MHz), the LWA (10--88 MHz), and the Allen
telescope (0.5--11 GHz). The correlated X-ray and radio exposures would
radically change the landscape of investigations for the disk-jet
connection for X-ray binaries and nearby active galactic nuclei\cite{fen06b}.
Partnerships would also be possible with new
wide-angle optical monitors, e.g., the Large Synoptic Survey Telescope
(LSST) and the Panoramic Survey Telescope and Rapid Response System
(Pan-STARRS). Prime targets would include tidal disruption flares
in nearby galaxies, a unique method to identify and measure the mass of
supermassive black holes in non-active galaxies.

Furthermore, the Sky Monitor will also serve as the electromagnetic ``eyes''
for associated events that can be detected by gravitational wave
observatories, such as Advanced LIGO and Advanced Virgo. As with all
types of survey detectors, there is substantial value in processing
data for a particular time and celestial position, compared to blind
searches with so many more statistical chances to find enhanced noise.
With its large field of view, the sky monitor could play a pivotal
role to establish the reality of particular events, while helping to
define the detailed science opportunities for gravitational wave
detectors.

\subsection{Mission Requirements}

\begin{table}[t]
\caption{Mission Requirements}
\small
%\begin{tabular}{llll}
\begin{tabular}{p{1.2in}p{0.75in}p{2.3in}p{1.8in}}
\hline\hline
Parameter &	Baseline & Drivers & Technology Factors \\
\hline\hline
    \multicolumn{4}{c}{\em Large Area Timing Array (LATA)}\\
\hline
Effective Area & 3.2 m$^2$& NS radius, BH QPOs  & Mass, cost, power \\
Minimum Energy & 1.8 keV	& Source states, absorption meas., soft srcs	& Detector electronics noise \\  
Maximum Energy & $>$30 keV	& BH QPOs, NS kHz QPOs, Cycl. lines & Silicon thickness \\ 
Deadtime	& 10\%@10~Crab$^*$ & Bright sources, X-ray bursts & Digital elec.\,design, pixel size  \\ 
Time Resolution & 1 $\mu$s & 	Resolve ms oscillations & Shaping time, GPS, Digital elec. \\
\hline
	\multicolumn{4}{c}{\em Sky Monitor (SM)} \\
\hline
Sensitivity (1~d) &	$<5$ mCrab$^*$ &	Faint transients, multi-source monitoring	& Camera size/weight/power\\ 
Sky Coverage &	$>2$ sr & 	TOO triggering, multi-source monitoring &	\# cameras vs. gimbaled designs\\ 
Source Location & 1 arcmin & Transient followup & Pixel size, camera dimensions \\
\hline
  \multicolumn{4}{c}{\em AXTAR Mission}\\
\hline
Solar Avoidance Ang.\hfill & 	30$^\circ$ &	Access to transients	& Thermal/Power design \\ 
Telemetry Rate & 1 Mbps &	Bright sources & Ground stations/TDRSS costs\\ 
Slew Rate & $>6^\circ$\,min$^{-1}$& Flexible scheduling, fast TOO response & Reaction wheels\\
\hline
\multicolumn{4}{l}{$^*$1 Crab = $3.2\times 10^{-8}$ erg~cm$^{-2}$~s$^{-1}$ (2--30 keV)}
\end{tabular}
\end{table}

The requirements set by our science objectives are summarized in
Table~1.  We emphasize the following points:
\begin{itemize}
\item{Our primary targets are bright Galactic X-ray sources.  Since
  the sky background is negligible in this regime, there is no
  advantage in concentrating optics, and a simple collimated detector
  will suffice. In particular, the detector must be able to handle the
  high photon count rates from bright black-hole transients,
  thermonuclear X-ray bursts, etc., corresponding to fluxes as high as
  10 Crab ($3.2\times 10^{-7}$ erg~cm$^{-2}$~s$^{-1}$, 2--30 keV), with
  minimal deadtime and pileup effects.  Furthermore, the data system
  must be able to store and transmit these data at high
  time-resolution and adequate spectral resolution without
  losses. Finally, the detector must have large effective area at
  energies well above 10~keV in order to have good sensitivity to the
  millisecond phenomena identified by RXTE.}
\item{Our required effective area is set by two key science
  objectives: (1) achieving a 5--10\% neutron star radius measurement
  using pulse shape modeling and pulse phase spectroscopy of
  thermonuclear X-ray bursts, and (2) achieving 0.1\% rms fractional
  amplitude detection threshold for high-frequency QPOs in accreting
  black holes.  Note that, for QPO signals 
  (i.e., resolved in frequency space), the signal-to-noise ratio of a
  given signal strength scales {\em linearly} with $A_\mathrm{eff}$, rather than the usual $A_\mathrm{eff}^{1/2}$ scaling\cite{vdk04}.}
\item{Our required lower energy bound is set by the need to determine
  temperature information from the thermal spectra of thermonuclear
  X-ray bursts, detect the soft thermal disk component of emission from
  accreting black holes, and measure interstellar photoelectric
  absorption toward our targets.  Our upper energy bound is set by the
  need for sensitivity to high-frequency QPOs in black holes and kHz
  QPOs in neutron stars, both of which are hard X-ray phenomena with
  maximum amplitude in the 10--30~keV range.}
\item{Many of our target observations need to be triggered to occur
  when the source is in a particular state; e.g., when an X-ray
  transient has emerged from quiescence into active outburst, or when
  an accreting neutron star is undergoing frequent thermonuclear X-ray
  bursts, or when an accreting black hole is in the particular
  spectral state where high-frequency oscillations are typically
  seen.  This requires that the mission include a sky monitor capable
  of triggering observations.  Since X-ray transients may appear
  anywhere in the sky at any time, and since there outbursts may be of
  short duration, this sky monitor should preferably cover the entire
  accessible sky (unocculted by the Earth and sufficiently separated
  from the Sun).}
\item{RXTE has shown that the events that trigger
  target-of-opportunity (TOO) observations may be very short-lived.
  It is thus essential that the mission can be quickly rescheduled in
  a matter of hours, and that it can slew to new coordinates
  expeditiously.}
\end{itemize}

\section{DETECTOR CONCEPTS}
\label{sec:detectors}  % \label{} allows reference to this section

For X-ray timing of bright sources, the key metric is effective area
at the relevant energies with low deadtime, while background rejection
and energy resolution are less critical than for some other
measurements. Therefore, focusing or concentrating optics are
generally best avoided in favor of large area detectors paired with
collimators to limit the field of view and reject the diffuse X-ray
background, which would otherwise be dominant.

Proportional counters have been the workhorse detectors for X-ray
timing measurements since the early days of X-ray astronomy. The state
of the art was moved forward by missions based on proportional
counters like EXOSAT (1983--1986, $A_\mathrm{eff}$ = 0.16 m$^2$ ),
Ginga (1987--1991, $A_\mathrm{eff}$ = 0.4 m$^2$),and most recently the
RXTE (1995--2010, $A_\mathrm{eff}$ = 0.6 m$^2$ ). However, the use of
proportional counters involves a number of challenges including modest
energy resolution, significant dead times, large mass and volume per
unit effective area, and a susceptibility to gas leaks and high-voltage
breakdowns.

Recently, a significant developments in solid-state detector
technologies have made possible a number of attractive replacements
for massive gas detectors. A recent example is the Large
Area Telescope (LAT) on NASA's \textit{Fermi} Gamma-ray Space
Telescope, which has a tracker that contains $\sim$75 m$^2$ of silicon
strip detectors rather than the gas spark chambers used in previous
experiments. The application specific integrated circuit (ASIC)
readouts for the strip detectors were rather simple, requiring a
fairly high energy threshold of $\sim 30$ keV and doing threshold
event detection only, not spectroscopy\cite{LATPaper}.

In the case of AXTAR, however, a low energy threshold ($\sim$2 keV)
and at least modest energy resolution are required.  Since most
solid-state detectors don't internally amplify the signal, and a 2 keV
X-ray liberates only $\sim$550 electron-hole pairs in silicon, very
low-noise pre-amplifiers are required to achieve these two goals. An
important component of the noise budget in such a readout is
proportional to the capacitance of the detector element and any
interconnect to the amplifier. This motivates a search for low
capacitance designs, and we briefly describe several alternatives here.

\subsection{Baseline concept: Si PIN diodes}

\begin{figure}
\begin{center}
\includegraphics[width=3.0in]{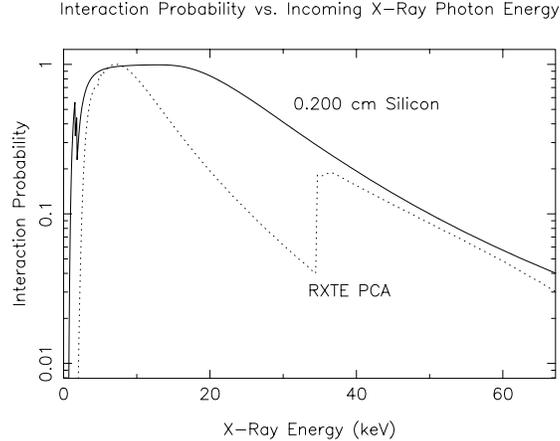}
\end{center}
\caption{Comparison of the photoelectric interaction probability for an X-ray photon
  in 2 mm of silicon with that in the RXTE PCA xenon proportional
  counters. The efficiency improvement in the 15--30 keV energy range
  is evident and means that the sensitivity improvement in that range
  is much larger than just the increase in geometric
  area. \label{fig:intprob}}
\end{figure}

Our baseline design, described in more detail below, is to use large
format (9$\times$9 cm active area) thick silicon detectors that are
divided into 2.5$\times$2.5 mm pixels, each of which is effectively
a simple PIN diode.  At 2mm thickness, the stopping power of silicon in the critical 15--30 keV band is much higher than the 1 atm xenon proportional counters used in the RXTE PCA (see Figure~\ref{fig:intprob}). The pixels are connected to readout ASICs via
either an interposer board that is bump bonded to the detector or by
wire bonds through a grid of holes in the interposer board. This
design is fairly straightforward and separates the design and
fabrication of the ASICs from the detectors.  One alternative is to
adopt a hybrid detector design where the ASIC is identical in size and
matched in pitch to the detector and the two are bump bonded
together. An advantage of this design is that the interconnect
capacitance can be made negligible compared to the pixel capacitance,
potentially lowering the power requirement to achieve the required
noise performance.  The disadvantages of the hybrid detector approach
may include low yields on very large ASICs and increased cost of
fabrication for the large area ASICs.  

\subsection{Alternative concept: Si drift detectors and CdTe}

An alternative is to replace the simple PIN diode pixels with a
more advanced detector configuration, such as a silicon drift detector
(SDD)\cite{gr84}.  An SDD uses carefully shaped electric fields to
direct the charge from a relatively large pixel to a very small
collecting anode, thus greatly reducing the effective capacitance of
the pixel. This results in significantly improved energy resolution
with reduced power requirements. However, SDDs are a newer technology
and are just now starting to be made in large area formats. Moreover, 
they have thus far been limited to fairly thin detectors 
($\sim$0.5 mm), so they may not have sufficient sensitivity above 10~keV for
our science requirements. 
One way to address that limitation would be a hybrid design,
with most of the area devoted to SDDs but a portion instrumented with
cadmium telluride (CdTe)\cite{oin+04} or another technology with good high-energy sensitivity.
An advantage of this hybrid approach is improved spectral fidelity at 
high energies relative to the thick-silicon option.  At 30~keV, Compton
scattering (and incomplete energy deposition) accounts for more than 10\%
of incident photon interactions in silicon but is a negligible effect in CdTe.
%(see Fig. 2, right panel).  
(Note that the xenon gas detectors in earlier 
missions also had negligible Compton scattering at these energies.)

All of these designs share several advantages of low dead time due to
the independent readout of each pixel and can provide precise timing,
energy, and pixel coordinates for each detected X-ray photon. For our
current concept study we have used the simple silicon pixel design as
a baseline. We will continue to study the alternatives and if one is
mature enough to be preferred, it will be substituted into the design.
From a mission design point of view this substitution will mainly
reduce the power requirement and improve the energy resolution of the
detectors.

\section{SCIENCE INSTRUMENTS}

\subsection{Large Area Timing Array (LATA)}

\begin{figure}
\begin{center}
\includegraphics[width=4.0in]{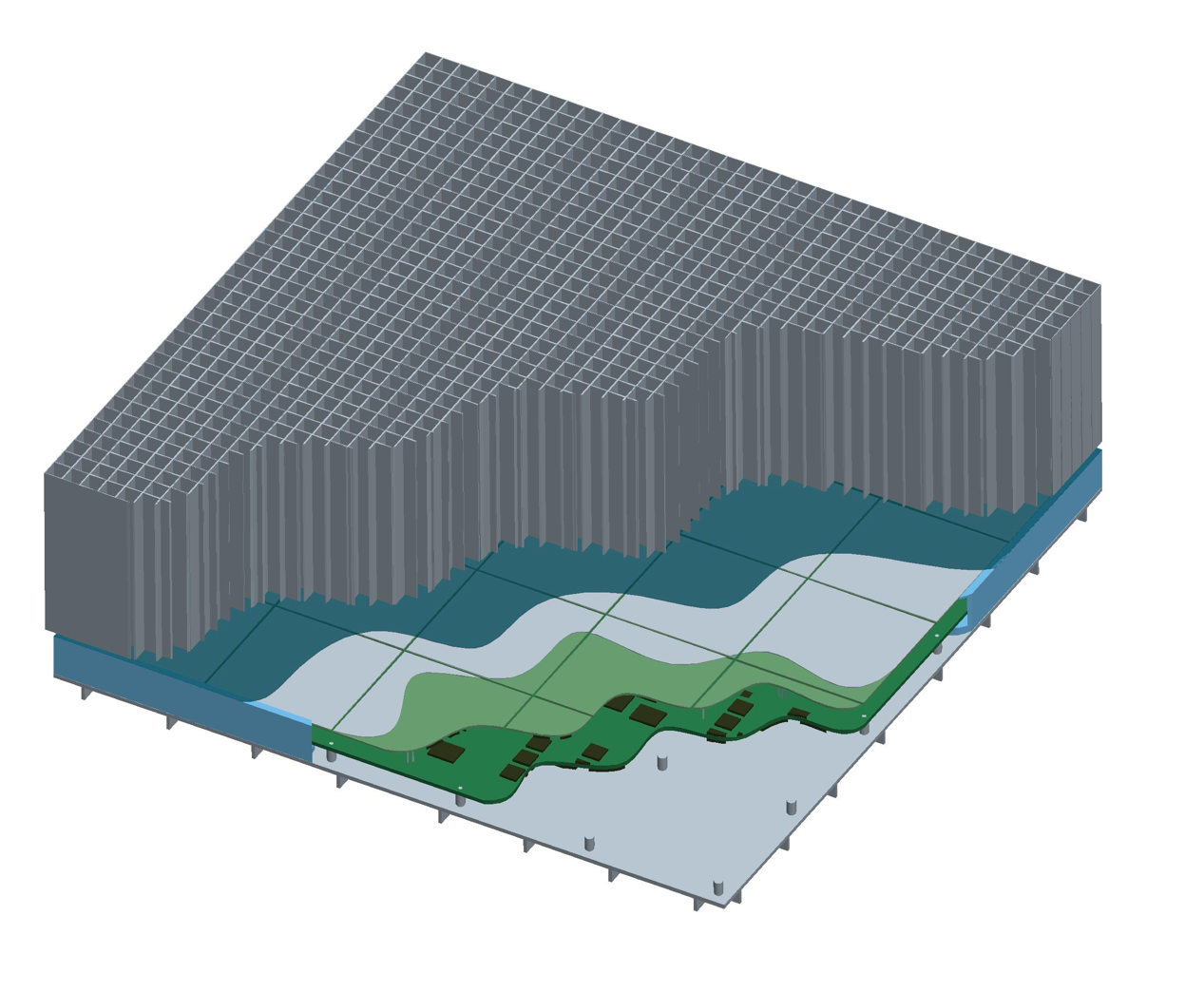}
\end{center}
\caption{Cutaway rendering of a LATA supermodule consisting of a 5 $\times$ 5 array of 10 $\times$ 10 cm detectors. The components from top to bottom are the collimator, light shield, silicon detectors, interposer board, and digital board, mounted in a box that provides support and shielding. Note that the collimator cell size is not to scale. \label{fig:supermodule}}
\end{figure}

The primary instrument on AXTAR is the Large Area Timing Array (LATA). For the LATA, we have chosen a modular design, consisting of an array of identical, but independent, detector `supermodules', each of which consists of a detector plane mated to a collimator and including support electronics (see Figure \ref{fig:supermodule}). This modular design ensures easy scalability to different size missions, robustness to individual failures, and simplifies assembly and testing.

\begin{figure}
\begin{center}
\includegraphics[width=3.0in]{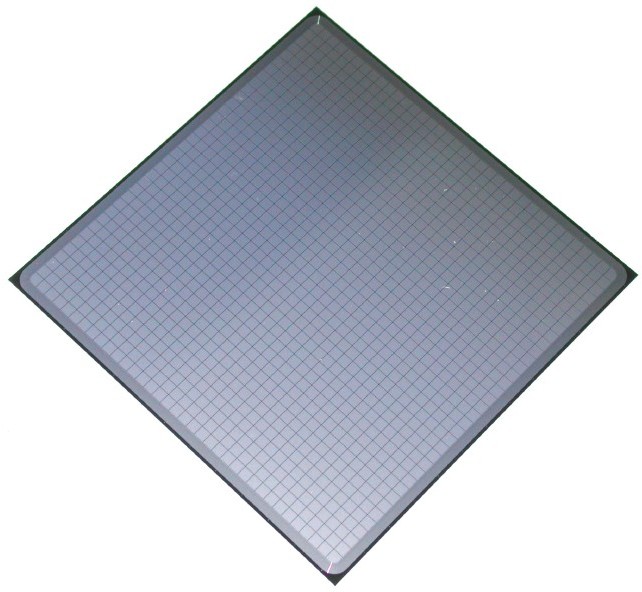}
\hspace{01.0in}
\includegraphics[width=2.0in]{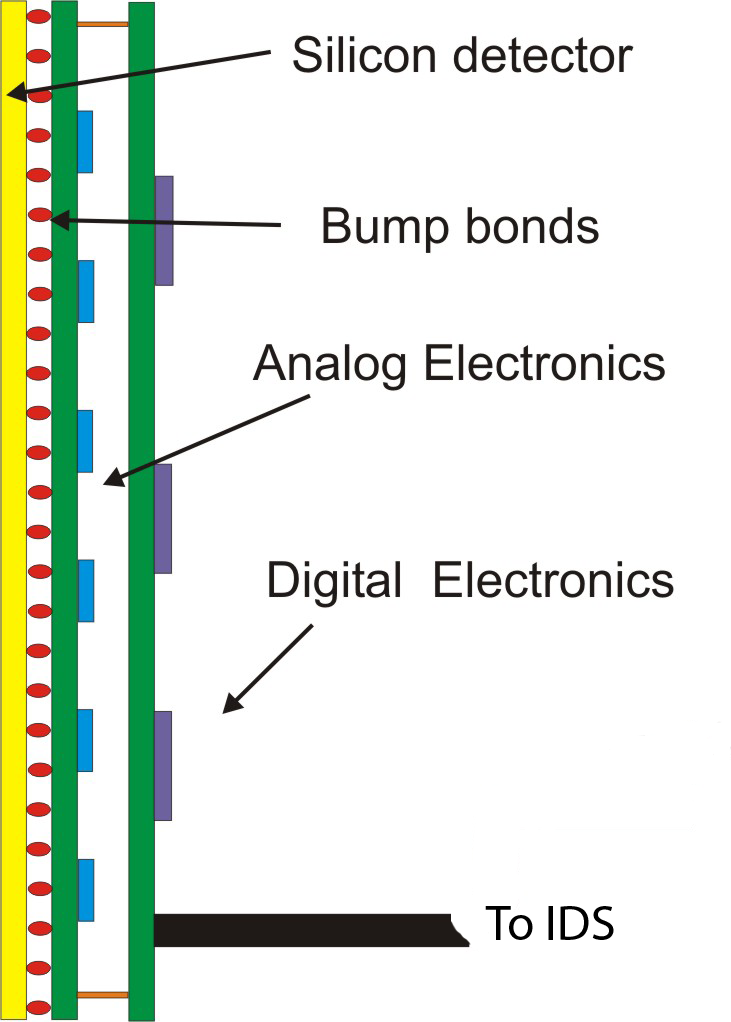}
\end{center}
\caption{\textbf{Left:} Photograph of a pixelated silicon detector for both the LATA and SM instruments with 36 $\times$ 36 pixels covering a total active area of 90 mm $\times$ 90 mm. \textbf{Right}: Side view of a LATA/SM detector assembly showing the layered construction of the detector module.\label{fig:detectors}}
\end{figure}

Each supermodule consists of a 5 $\times$ 5  array of detectors, front end electronics, a digital interface to the Instrument Data System (IDS), and low and high voltage power supplies.
The baseline detectors (see Figure \ref{fig:detectors}) are manufactured using 150 mm diameter high-resistivity wafers that are 2.0 mm thick. The detectors are 96 mm $\times$ 96 mm in size with an active area of 90 mm $\times$ 90 mm and with a 3 mm wide guard structure around the perimeter of the detector. Each detector is segmented into a two-dimensional array of 36 $\times$ 36 pixels, each with an area of 2.5 $\times$ 2.5 mm$^2$. This segmentation was selected to minimize the noise of the analog read-out electronics for a given power budget, in this case 1 W per detector. A single contact that is biased at high voltage ($\sim 700$V for this thickness) covers one side of each detector; this is the side exposed to the incoming X-radiation.  Each supermodule thus provides 2025 cm$^2$ of total active area, which becomes 1720 cm$^2$ of active area for a collimator open fraction of 0.85.

In the baseline design, the silicon detectors are bump bonded to an interposer board that provides mechanical support and connections to the critical front-end electronics. The interposer board is manufactured from very low dielectric loss material and the traces are designed to minimize stray capacitance from the connection. To keep the trace lengths low, the readout ASICs have 36 input channels and read out a 6 $\times$ 6 array of pixels. A total of 36 ASICs are required to read out all 1296 pixels on the detector. Preliminary modeling shows that using Rogers 5870/5880 material (with a dielectric constant of 2.4/2.33), the capacitance per pixel will be between 0.3 pF and 0.6
pF, depending on the pixel position.  If more than 36 pixels
are connected to a single ASIC, the loss of performance due to the
capacitance from the traces outweighs the power savings and the
overall performance is degraded.

The critical readout ASIC must be designed to be both low power and low noise to achieve the required low energy threshold of 2 keV.  To achieve an acceptable limit on noise-induced counts in the electronics, we compute that the equivalent noise charge (ENC) of the of the amplifier must be $<77$ electrons r.m.s.\ We thus set the design goal for the front end ASIC at 70 electrons r.m.s.\ to provide some margin.  With this noise, the detector will
have an energy resolution of 600 eV FWHM.  This energy resolution would be nearly constant over the full energy range of the detector. The BNL group has designed a similar ASIC\cite{dcf+07} that meets the AXTAR performance goals by more than a factor of two, achieving an energy resolution of 173 eV at 6 keV, but consumes a factor of two more power (1.6 mW/channel) than our budget.  Development of a new CMOS ASIC design with lower power is planned. The functions performed by the ASIC include charge amplification, shaping,  threshold discrimination, peak detection, and multiplexing of the output to the digital board.  The ASIC will also include circuitry to adjust gain settings, inject calibration pulses, mask off bad channels, and set discriminator threshold levels.

All the digital electronics for a module will be located on a second board
with the same form factor as the detector and interposer board.  This board
will contain the analog-to-digital converter (ADC) and a Field Programmable
Gate Array (FPGA).  The FPGA will control the detector ASICs and the ADCs.
The digital electronics board will produce a stream of time-tagged events that
will be sent to the Instrument Data System (see \S\ref{sec:ids}).

Our baseline collimator concept is based on the design used for the RXTE PCA \cite{jmr+06}. There will be one 50 $\times$ 50 $\times$ 20 cm$^3$ collimator per supermodule, with a nominal field of view of 1$^\circ$ (FWHM).  Each collimator module is manufactured from Be-Cu foils. Each collimator will support a thin film thermal shield that is largely transparent at $E > 2$ keV to the incoming X-radiation. Because of the small mass of the silicon detectors, the mass budget of the instrument is dominated by the collimators. We are continuing to study alternative collimator designs that could reduce the instrument mass without significantly degrading the performance.  In addition to the collimator, there will be a graded-Z (e.g. Ta-Sn-Al or Ta-Cu-Al similar to what was used for the RXTE PCA\cite{jmr+06} and the Swift BAT\cite{rob03}) shield on the back and sides of the instrument, primarily to block diffuse cosmic X-rays from hitting the detector. Because the Sun becomes an exceedingly bright source of soft X-rays during solar flares, we also include sun shields (nominally 0.2 mm thick lead foils) that keep the Sun from directly illuminating the front of the collimators for all pointing directions $>45^\circ$ from the Sun. This prevents singly-scattered solar X-rays from being a significant background source during periods of high solar activity.

Each complete supermodule has an estimated mass of 30 kg and a power budget of 30 W. For the electronics to maintain their noise performance, they must be maintained at a temperature of 10$^\circ$C or lower.

\begin{figure}
\begin{center}
\includegraphics[width=3.0in]{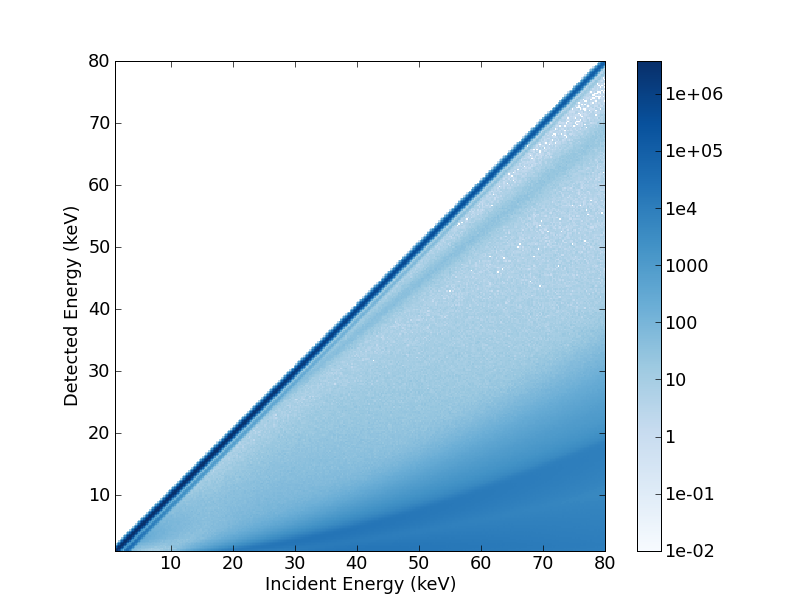}
\includegraphics[width=3.0in]{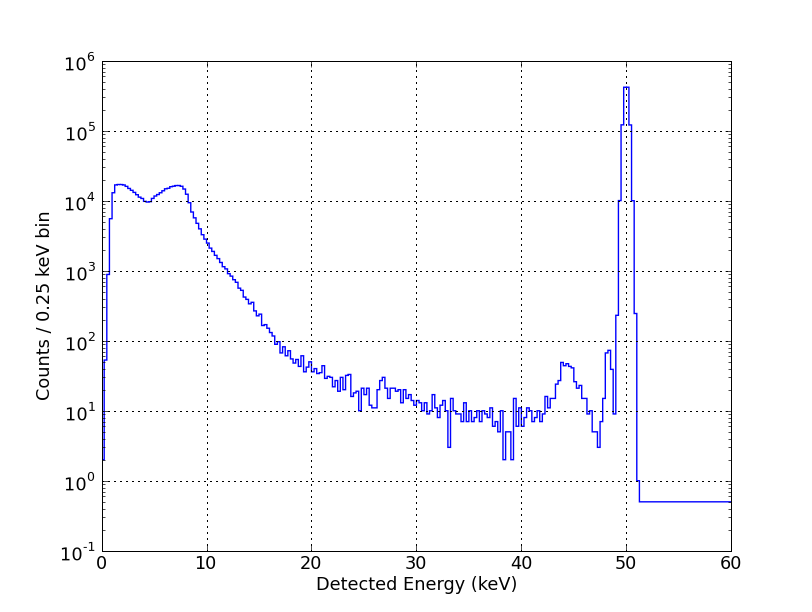}
\end{center}
\caption{Response matrix simulation results for a baseline 2-mm thick LATA silicon detector. The panel at left shows the 2-dimensional response matrix with detected energy as a function of incident photon energy, with a log color scale. The effects of escape of fluorescence photons and Compton scattering are responsible for the non-diagonal terms. At right is the response for monochromatic 50 keV incident photons.\label{fig:rsp}}
\end{figure}

We have modeled the AXTAR LATA spectral response using the \texttt{Geant4} particle simulation program \cite{geant4}. In these simulations we illuminated one LATA module of 9 cm by 9 cm silicon (2 mm thick) with a monochromatic beams of 10$^7$ X-ray photons oriented normal to the detector face spaced at 0.25 keV intervals covering the range 1.25 to 80 keV. The simulations include a detailed treatment of many radiative effects including photoelectric absorption, fluorescence, and Compton scattering of the injected photons. Our primary interest here is to understand the effects of Compton scattering on the spectral response of the LATA. Figure \ref{fig:rsp}  shows the simulated response of the LATA module. The left panel is a summary of the resulting detected energies from illuminating the silicon. Photoelectric absorption of X-rays leads to the primary photopeak and associated escape peak at the energy of the injected photons. However, Compton scattering results in a continuum of detected photon energies at lower energies within the LATA energy range. The detailed redistribution of 50 keV photons by the active silicon is shown in the right hand panel. The spectrum of detection energies from each discrete injection energy can be utilized to create an energy redistribution matrix for each input X-ray energy. This redistribution function is then incorporated into a response matrix suitable for simulation of the LATA response to astrophysical source spectra.

The instrument must reject several types of backgrounds, most importantly from charged particles in the space environment.  Low energy charged particles will be stopped by the same shielding that blocks diffuse X-rays but shielding against high energy particles can never be completely effective.
The passage of high-energy charged particles through the detector will almost always produce ionization that is equivalent to a photon of energy of $>50$ keV or more; i.e., well above the sensitive band of the instrument, and can therefore be rejected by an energy cut on the detected signal. Particles that ``nick a corner'' of a pixel will need to be identified on the basis of a coincidence with a high energy event in a neighboring pixel. In any case, there are a number of means of identifying and rejecting all types of non-X-ray-induced events with efficiencies that should be adequate.

\subsection{Sky Monitor (SM)}

The AXTAR-SM is designed to independently pursue the science themes of
wide-angle X-ray astronomy (\S\ref{sec:sci}), which includes the
detection of target opportunities for AXTAR pointed observations.
However, the SM will not drive a major increase in the overall cost of
the mission, as it is constrained to require no more than $\sim$20\%
of the resources required for the LATA, i.e., for cost, mass, volume,
downlink rate, and power. The design is greatly facilitated by using
precisely the same detector technology developed for the main
detector array and by avoiding the use of moving parts. The SM
operations plan should not impose any constraints on the LATA
observing program.  Furthermore, the SM cameras must be able to be
pointed anywhere in the sky, including fields containing the Sun,
although no useful data will be produced from a camera while the Sun
in its field of view.

\begin{figure}
\begin{center}
\includegraphics[width=2.5in]{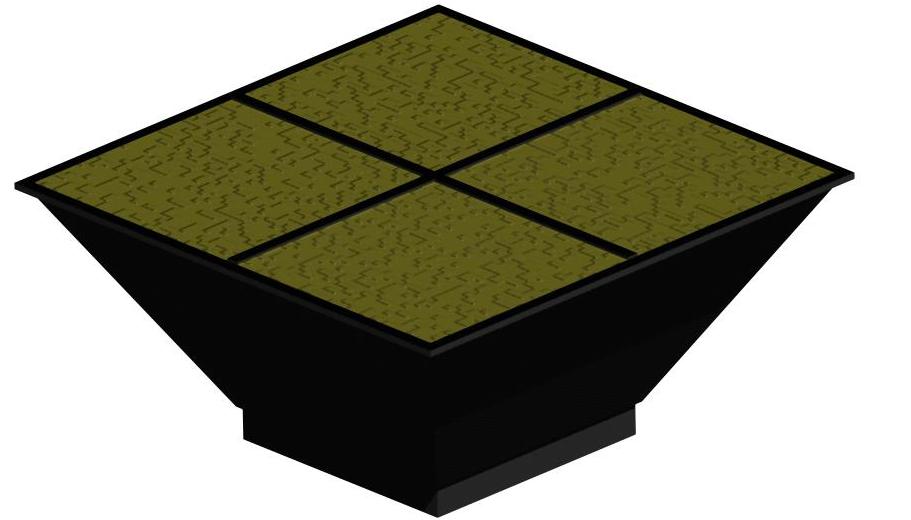}
\hspace{0.5in}
\includegraphics[width=2.5in]{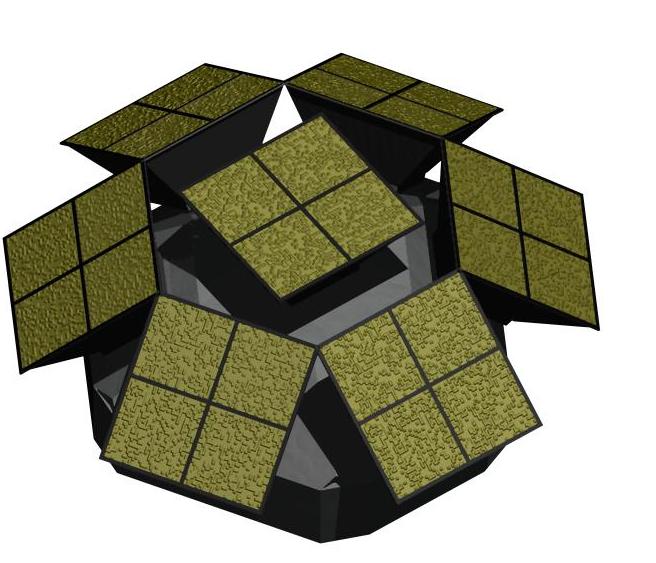}
\end{center}
\caption{\textbf{Left:} Rendering of a Sky Monitor camera. \textbf{Right:} A cluster of 7 SM cameras.\label{fig:sm}}
\end{figure}

The baseline design for the SM is a set of coded-mask cameras with
detector planes consisting of a 2 $\times$ 2 array (300 cm$^2$ total
effective area) of the same Si pixel detectors (2.5 mm pitch) used for
the LATA. Each 2 $\times$ 2 detector array will be mounted in a camera
body topped with a 2-D coded aperture mask (see Figure \ref{fig:sm}). Each camera will be approximately 25 cm square at the base
(detector end) and 30 cm square at the aperture plane.  The total
height will be approximately 45 cm. The cameras are relatively light;
each will have a mass of a few kilograms, at most.

The field of view of each camera will be approximately 40$^\circ$
$\times$ 40$^\circ$ (FWHM) or about 0.4 sr. For this size field, full
sky coverage requires a complement of 32 cameras.  The concept that
was studied in detail in the recent exercise comprised 27 cameras,
grouped in 5 clusters. This instrument would achieve $\sim 85$ \%
all-sky coverage (to FWHM), with most of that coverage having uniform
effective area. Further coverage losses are expected from Earth
occultation (33\%), passages through the South Atlantic Anomaly (an
additional 7.5\% at $28^\circ$ orbit inclination, and 2\% at
$5^\circ$), and AXTAR slews (few \%). In the end, the AXTAR-SM can
achieve an average of 50--55\% in all-sky live-time coverage, which
far exceeds the levels of previous missions (see \S\ref{sec:sci}).

\begin{figure}
\begin{center}
\includegraphics[width=3.5in]{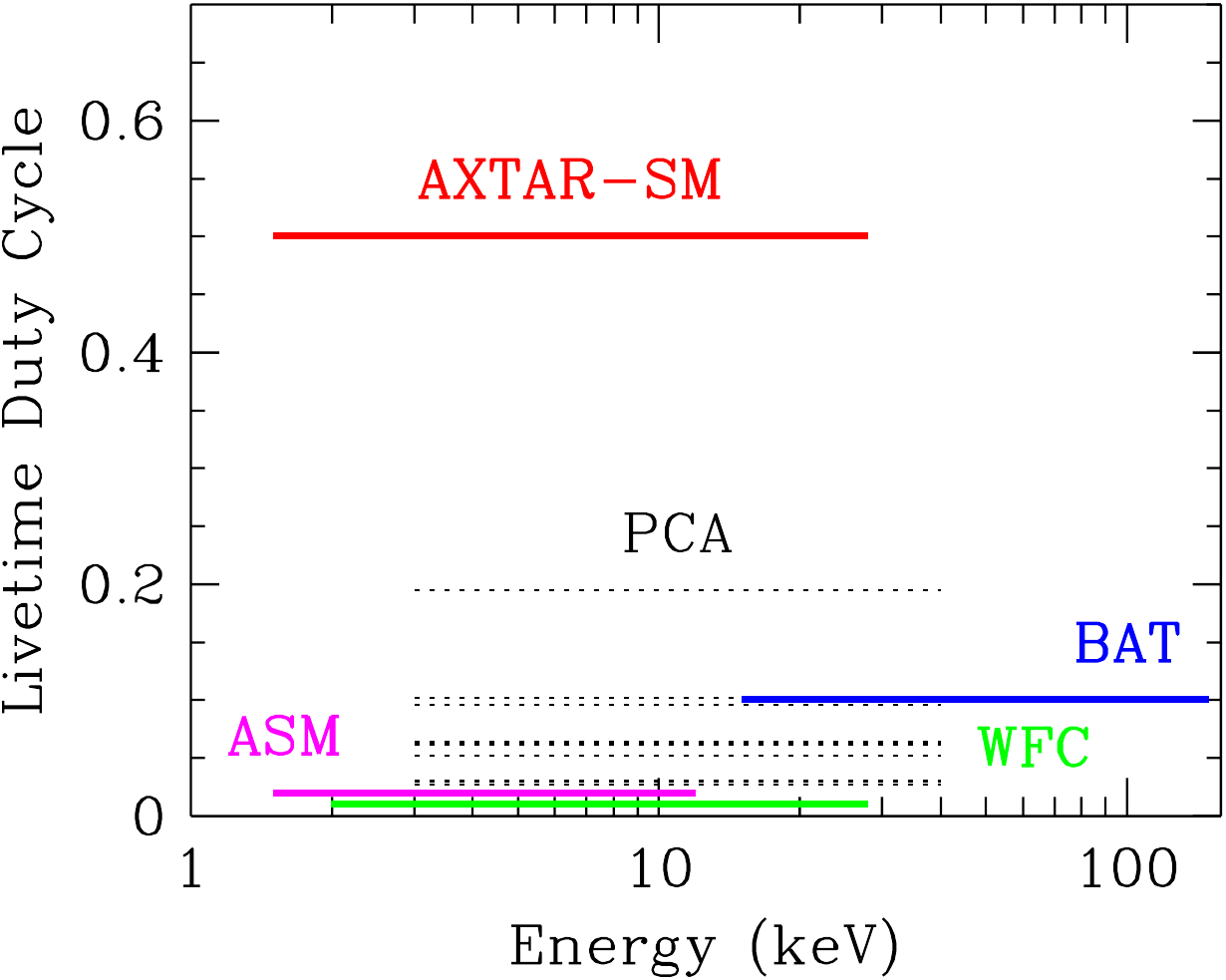}
\end{center}
\caption{The livetime duty cycle of various X-ray instruments plotted versus
photon energy.  Solid, colored lines show all-sky livetime averages
($\tau$) from wide-angle X-ray monitors.  The dotted lines show the
highest duty cycles achieved for individual sources during monitoring
campaigns of several days with the RXTE PCA instrument.  The AXTAR-SM,
in the 27 camera case, can achieve $\tau = 0.5$, or higher, while its bandwidth (2--30 keV)
is sensitive to both the thermal and non-thermal components radiated
by accreting black holes and neutron stars.\label{fig:livetime}}
\end{figure}

Thus far, the best wide-angle X-ray monitors have achieved duty cycles
per source: $\tau = 0.020 \pm 0.005$, depending on ecliptic latitude,
for the RXTE All Sky Monitor (ASM), and $\tau = 0.010$ for the 
Beppo-SAX Wide Field Camera (WFC). For bright X-ray transients, RXTE
pointed observations are typically conducted with $\tau = 0.025$,
while a singular large-program experiment for the 2005 outburst of
GRO~J1655-40, achieved $\tau = 0.096$.  These values are insufficient
to capture critical, infrequent events that convey the greatest
leverage for constructing detailed physical models.  The Swift BAT
(1.4 sr FOV) is producing hard X-ray light curves (15--50 keV) for XRBs
with $\tau \sim 0.10$.  These data are extremely useful in many ways,
but for the investigations of accretion disks, state transitions, and
the disk:jet connection, there is the limitation that the BAT
sensitivity is above the spectral range of both black hole and neutron star disks
(typically 0.5--1.5 keV at the inner boundary) and also the neutron star boundary
layer (blackbody $T < 2.7$ keV).

In the current AXTAR design, the capability of the SM to achieve $\tau
= 0.5$ is a historical step forward, as illustrated in Figure \ref{fig:livetime}. This
provides data at the times critical interest for diverse science
topics outlined in \S\ref{sec:sci}.  Furthermore, the one-day sensitivity would
be improved by almost an order of magnitude, compared to the RXTE
ASM, reaching 1 mCrab at $4 \sigma$, or $1\times~10^{35}$ erg s$^{-1}$
at the Galactic center. An archive of good events from the 
SM with excellent time resolution (e.g., 122 $\mu$s) would provide
retrospective capability to process data for any position, at any time
of interest.

\subsection{Instrument Data System (IDS)}
\label{sec:ids}
The LATA will generate 120 kcts s$^{-1}$ from the Crab, 0.5 Mcts s$^{-1}$ from a 4
Crab black hole transient, and 1.2 Mcts s$^{-1}$ from Sco X-1! An event-by-event mode
with, e.g., 40 bits per event would require a 5 Mbps data stream during
observations of a 1 Crab source. Therefore adequate data handling
capability is crucial to maintain event throughput and to optimize the
usefulness of the data products that are chosen for transmission
within the limit of telemetry bandwidth.

RXTE had great success observing bright sources by using a highly
flexible Experiment Data System (EDS) with programmable data modes
that could maximally utilize the telemetry bandwidth available. We
propose a similar data system for AXTAR, the Instrument Data System
(IDS). The IDS will, like the the EDS, generate multiple data products
for the LATA by using event handlers, in parallel, to implement each
chosen data ``mode''. This will allow, e.g., capture of one data
stream with the full energy resolution of the detectors at low time
resolution, while also sending down data at very high time resolution
with modest energy resolution in another stream. If the number of
counts per time or energy bin is much greater than unity (as it would be for bright
sources or low time resolution), the data will be binned. If it is much less than unity, then the telemetry would be minimized by transmitting the
elemental data for each photon event. The IDS will be fully
reprogrammable to allow it to adapt to new ideas, new discoveries, and
other unexpected conditions.  On the other hand, we expect that one or
two ``standard modes'' will always be operating to create a uniform
data set to facilitate archival analyses.

\section{MISSION DESIGN}
\label{sec:mission}  % \label{} allows reference to this section

In this section, we describe the baseline mission design, developed as part of a mission concept study at the MSFC Advanced Concepts Office.  RXTE, launched in 1995, cost more than \$200M. In our evaluation, it is not feasible to build a successor mission with a factor of several more geometric and effective area within the confines of a NASA SMEX (the last call was in 2008 for \$105M, not including launch) or EX class mission (anticipated 2010 for \$200M, not including launch). However, with the cost and mass savings possible using large area solid state detectors, we believe that a major improvement over RXTE can be made with a MIDEX class mission.  For planning purposes, we hypothesize a 2014 call for proposals for a $\sim$\$300M (not including launch) class mission to be launched in 2019.

\subsection{Assumptions and Requirements}

The spacecraft was designed to meet the requirements listed in Table \ref{tab:assumptions}. As the requirements are relatively modest, no new technologies are needed for the spacecraft to enable the science mission.  The Table also include key data on the science instruments that affect mission design.

\begin{table}[t]
\caption{AXTAR mission study parameters\label{tab:assumptions}}
\small
\begin{center}
\begin{tabular}{ll}
%\begin{tabular}{p{1.2in}p{0.75in}p{2.5in}p{1.8in}}
\hline\hline
Parameter & Required Value (Desired or Maximum Value) \\
\hline\hline
\multicolumn{2}{c}{AXTAR spacecraft} \\
\hline
Orbit Altitude & $\sim 600$ km (study output), circular \\
Orbit Inclination & 28.5$^\circ$ or less (as low as possible) \\
Spacecraft Lifetime & 3 yr\\
Consumables & $>5$ y\\
Orbit Lifetime & 10--15 yr \\
Pointing Accuracy &	$<1$ arcminute \\
Pointing Knowledge & $<5$ arcsec  \\
Maximum Slew Rate & 180 deg in 30 minutes  \\
Maximum Continuous Observing Time & 28 hr \\
\hline
\multicolumn{2}{c}{AXTAR Sky Monitor camera (each)} \\
\hline
Mass &   2 kg + 2 kg per telemetry hub  \\
Power &   4W + 9W per telemetry hub    \\
Quantity & 7 minimum  (32 maximum) \\
Thermal requirement & $-40^\circ$C to +10$^\circ$C (detector plane) \\
Alignment    &   32 faces and vertices of dodecahedron    \\
\hline
\multicolumn{2}{c}{AXTAR LATA supermodule (each)} \\
\hline
Mass & 30 kg \\
Power & 30W  \\
Quantity & 20 (as many as possible) \\
Thermal requirement & $-40^\circ$C to +10$^\circ$C (detector plane) \\
Alignment & all co-aligned within 1 arcminute \\
\hline \hline
\multicolumn{2}{c}{Contingency Philosophy} \\
\hline
Mass & 30\% for  spacecraft subsystems and instruments \\
Power & 30\% for spacecraft subsystems and instruments \\
\hline
\end{tabular}
\end{center}
\end{table}

\subsection{Mission Analysis}

The orbit selection was driven primarily by three criteria: to minimize passage through the South Atlantic Anomaly (SAA), to avoid radiation at the higher altitudes, and to avoid as much atmospheric drag as possible. While lower inclinations help with avoiding the SAA, the science instruments can be turned off during these periods, so this is not a driving requirement. In order to avoid higher radiation exposures, the altitude must be limited to not much more than 600km. And finally, avoiding atmospheric drag requires a higher altitude orbit, preferably one in which natural orbital decay will not result in re-entry sooner than 10 years after orbit insertion, which minimizes station-keeping. Using the NASA Debris Assessment Software (DAS 2.0)\cite{das} and Analytical Graphics Satellite Took Kit (STK)\footnote{\url{http://www.agi.com}}, the team determined through parametric analysis that the optimal initial orbit would be a 585 km circular orbit, with the goal for the launch vehicle to insert the observatory into as low of an inclination as possible. To be slightly conservative, the initial orbit altitude for use in determining launch vehicle performance was set to 600 km.

\begin{figure}
\begin{center}
\includegraphics[width=6in]{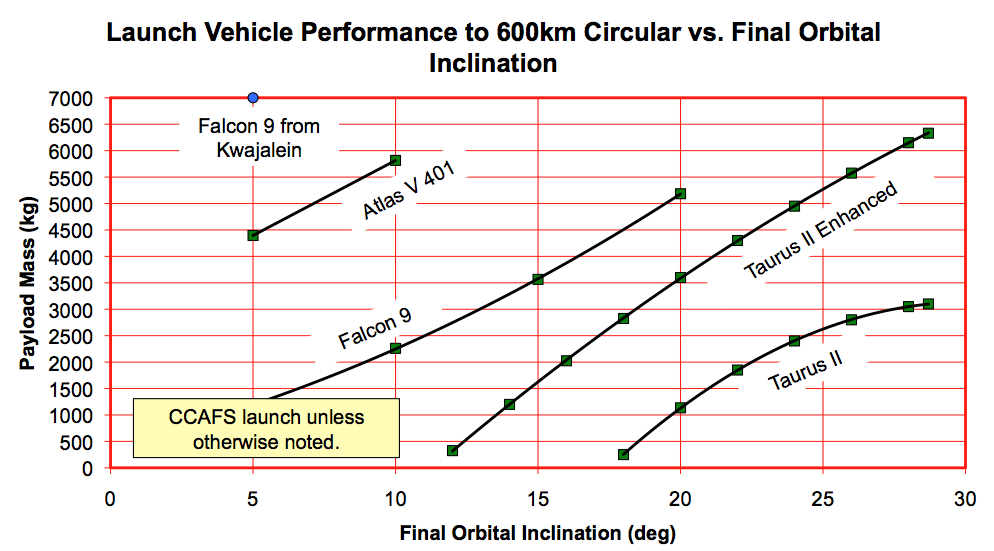}
\end{center}
\caption{Payload mass versus inclination for various launch vehicles.}\label{fig:payload_v_inclination}
\end{figure}

Using historical spacecraft data for other science missions, the team estimated the observatory mass would be approximately 2000kg for a configuration with 20 LATA supermodules. Given this mass, and the estimated area required by 20 LATA supermodules, the team chose two launch vehicle candidates: the Orbital Sciences Corporation's Taurus II, and the SpaceX Falcon 9 (which had a successful inaugural flight on June 4, 2010). Since both vehicles can use the same payload adapter, a spacecraft that will fit within the Taurus II shroud will also fit within the Falcon 9 shroud, giving at least two vehicle options for the spacecraft. Performance estimates for the Taurus II came directly from Orbital Sciences Corporation\cite{taurusII}; for the Falcon 9, the data were provided by NASA Launch Services Program\footnote{\url{http://elvperf.ksc.nasa.gov/elvMap/}}. Since the goal is to achieve as low an inclination as possible, any excess payload capability provided by the launch vehicle will be used to lower the orbital inclination. The payload maximum mass versus inclination for various launch vehicles is plotted in Figure \ref{fig:payload_v_inclination}. Note that launching the Falcon 9 from Kwajalein results in the most mass to the lowest inclination, but may incur additional costs. The Taurus II Enhanced data are very preliminary, as this vehicle is still in the design phase, and is offered only for comparison. 

\subsection{Spacecraft Configuration for Taurus II Class Launch Vehicle}

The driving factor in the spacecraft's configuration was the launch vehicle size which limited the number of primary science instruments (LATAs). The LATAs coplanar and pointing requirements drove the design layout to be divided into two main sections, a spacecraft system bus and a science bus.  The science bus/LATA array was set forward to provide separation from the spacecraft's solar arrays and other systems. The spacecraft system bus is located aft near the launch vehicle interface to carry launch vehicle loads and to allow for the spacecraft's systems to be enclosed in a more compact volume. 

The other design factor was the number and placement of the  sky monitor cameras.  A configuration using the 32 combined faces and vertices of a dodecahedron would allow full sky coverage but due to volume and operational constraints 27 were chosen for the baseline configuration. These were broken up into several clusters and placed on the spacecraft to optimize viewing angles while minimizing size. 
The overall configuration is conservative and does allow room for component growth and for extra subsystem components to be added that were not analyzed in this study.  The design has continuous load paths and structures which allows for minimum mass to be obtained.  The design, while based on the Taurus II launch vehicle, could be used on comparable (not smaller) size launch vehicles.  If a larger launch vehicle was selected a different overall configuration and layout would need to be studied to achieve the optimal configuration. 

The design includes a 30\% mass and power contingency for all spacecraft systems and for all science instruments. The spacecraft design for the Taurus II resulted in a total vehicle gross mass of 2,650 kg, which is outlined in the table below. Gross mass is the combined total of vehicle dry mass, inert mass, and propellant.  Dry mass is defined as spacecraft subsystems mass minus the useable propellant, propellant residuals, and science instruments. Inert mass includes propellant residuals and science instruments.

\begin{table}
\caption{Master equipment list and mass budget (Taurus II configuration)\label{t2MEL}}
\small
\begin{center}
\begin{tabular}{lr}
\hline\hline
Equipment  & Mass (kg) \\
\hline\hline
Structures (incl.\ LATA radiation shielding) &  745 \\
Propulsion & 66 \\
Power & 137 \\
Avionics, Control, Comm & 189 \\
Thermal Control & 39 \\
Contingency (30\%) & 351 \\
\hline
DRY MASS & 1527 \\
\\
Non-propellant fluids & 4 \\
LATA (20) supermodules & 600 \\
Sky Monitors (27) & 54 \\
Instrument Data System & 20 \\
Payload contingency (30\%) & 202 \\
Instrument cabling & 35 \\
\hline
INERT MASS & 915 \\
\\
TOTAL LESS PROPELLANT & 2442 \\
\\
Propellant (Hydrazine) & 208 \\
\hline
GROSS MASS & 2650 \\
\hline
\end{tabular}
\end{center}
\end{table}

\subsubsection{Spacecraft Structure}

The octagonal AXTAR bus provides a launch vehicle adapter and structure to mount all instruments and 27 sky monitors providing nearly full sky coverage, as well as, maintaining 20 LATA supermodules in a coplanar array.  Radiation shielding against cosmic and solar radiation is incorporated into the spacecraft bus, and is considered as part of the spacecraft structural mass in Table \ref{t2MEL}.  The spacecraft utilizes lightweight 2024-T351 aluminum panels, tubing struts, and frames for component mounting and to double as radiators for thermal management. Two flight-proven telescoping solar array booms\cite{mps99} are stowed against the spacecraft bus for launch.  A load set of combined Taurus II/Falcon 9 loads, 6.0 g axial and 2 g lateral, was applied at 30 and 45 degrees around the launch axis. Finite Element Modeling Analysis and Post-processing (FEMAP) verified a positive Margin of Safety for a spacecraft structure loaded with instrument masses and analyzed with NX NASTRAN using a 1.4 Factor of Safety for isotopic strength.

\subsubsection{Communications System}

At present, the downlink data rates for this mission are within the capabilities of a system using a fixed antenna.  A communication link to ground using fixed antennas is desirable over an active pointing design because it eliminates gimbaled mechanisms, and using an omni-directional antenna is more reliable and reduces mission risk. Analysis shows that using an X-band system with omni antennas for the science data downlink is sufficient for the given data rates and assumed mission scenarios.  Using TDRSS for normal operations and data link is an alternative design that was not pursued in this study.  However, a TDRSS link during launch and start-up operations is desirable.  The communication system designs are single fault tolerant.

For this mission, low orbit inclinations are better for science data collection.  A ground link analysis based on link times and daily accesses was performed to determine the best selection of ground stations at both 5 and 28 degrees.  Out of eleven possible ground stations analyzed, Southpoint Hawaii and Kourou Guiana were selected as the primary and secondary ground link stations respectively for 28 degree orbits.  These same two stations are also selected at a 5 degree orbit, where Kourou becomes the primary and Southpoint becomes the secondary stations.

At these two stations, at least an 8 minute primary link is possible 7 times a day.  Under a worst case condition, a 6.7 minute link 5 times a day is possible for the 5 degree inclination at the secondary Southpoint station.  Using these two conditions as bounds, a link budget was performed assuming at least 4 links per day for 8 minutes each.  For these link times, it was found that a 20 watt transmitter can achieves a 90 Mbps data downlink rate.  At this rate, almost 44 Gbits can be transmitted per station pass, or 172 Gbits per day.  Assuming a continuous average data rate of 700 kbps for the entire LATA system and 72 kbps for an 8 unit SM cluster, enough link capability is left to download over six 15 minute peak (19 Mbps) LATA events per day.  Major SM events can be alternated with LATA event data, assuming the SM detects events first and then the LATA is pointed to the event.  The suggested 20 watt L3 X-band transmitter is presently at TRL\footnote{Technology Readiness Level, see \url{http://esto.nasa.gov/files/TRL_definitions.pdf}} 6.

An estimate for the total spacecraft telemetry communication data rate using S-band is 60 kbps downlink and 4kbps uplink.  Using a 5 watt S-band transmitter, a link to TDRSS for launch and start up operations can be accomplished at these data rates, again using omni antennas.  The 5 watt transmitter can also easily link with ground for normal telemetry with plenty of margin available.  The suggested AeroAstro 5 watt S-band transmitter is at TRL 8.

\subsubsection{Power Systems}

The overall power demand, including the spacecraft, science instrumentation and 30 percent growth margin, is 1583W for the Taurus II configuration. The Power System supplies all of this demand. Power is generated by conventional, rigid panel solar arrays configured from space-qualified GaAs 3-Junction cells. The solar array regulation, power conversion, and power distribution is performed by a set of VME power electronics boards designed for use on the Mars Reconnaissance Orbiter and Orion spacecraft. All power function circuits are redundant and the boards are hosted in a single VME enclosure with redundant power supplies. The bus voltage is 28V. Energy storage is provided by Lithium Ion batteries configured from existing, space qualified cells. The maximum depth of discharge allowed is 40 percent. One `Hot Spare' battery is carried for redundancy. The power system mass (including 30 percent contingency) is 179 kg for the Taurus II configuration.

\subsubsection{Avionics and GN\&C}

Two fully redundant Proton 200 flight computers from SpaceMicro are the core of the avionics system.  The computers are used for spacecraft operations and data management.  They receive processed science data from the IDS, and transfer the data to either the on board data recorders or the X-band transmitters for downloading.  Two data recorders from Surry Satellite Technology can store up to 256 Gbits of data at a rate of 150Mbps.  One day of required data storage is estimated to be about 173 Gbits, giving approximately a 50\% memory margin.  The flight computers are radiation hardened to 100 krad total ionizing dose and 70 MeV-cm$^{2}$/mg single event latch-up. The Proton 200 is scheduled for its first launch in 2010, and is presently at TRL 6.  The Surrey data recorders are at TRL 8.

Attitude knowledge  is achieved using a redundant pair of Ball Aerospace star trackers and Northrop Grumman IMUs.  The star trackers provide 4 arcseconds of accuracy, meeting the 5 arcsecond mission requirement.  Both the IMUs and the star trackers are at or above TRL 8.

While this satellite has large surface areas resulting in significant disturbance torques, the slewing and pointing requirements are modest.  Off-the-shelf reaction wheels should be sufficient for attitude control, keeping cost down.  In low earth orbit, magnetic torque rods are good candidates for attitude control assistance and reaction wheel desaturation.  Using magnetic torquers, the spacecraft reaction control system will not normally be required for attitude control, and is considered for contingency purposes and disposal only.  A set of 3 dual coil Microcosm MT400-2 magnetic torquers is suggested for this mission.  These torquers are at or above TRL 8.

The trade space for reaction wheels included 4 different wheel types at four slewing speeds.  This trade resulted in a Teldix RSI 68-170 reaction wheel being selected.  Mounted in a 4 wheel pyramid configuration for best performance while maintaining one fault tolerance, these wheels have sufficient momentum and torque capability to exceed the fast slew requirement of 180 degrees in 30 minutes, while providing the desired pointing accuracy of 1 arcminute.  The Teldix wheels are at or above TRL 8.

Inertial pointing will be the typical pointing mode for science observations.  In this mode, the magnetic torque rods provide good torque authority.  Using the torquers to offset the disturbances, the 28 hour continuous pointing goal can be achieved, with unlimited pointing times possible.  In a zenith pointing mode, a possible scanning mode, the disturbances will usually be small.  In this pointing mode desaturation of the reaction wheels can be accomplished using the magnetic torque rods.  In a worst case torque scenario the atmospheric and gravity torques are at a continuous maximum.  If left in this mode for extended periods the reaction wheels will eventually saturate.  However, extended pointing times in this mode are unlikely since there are no apparent science advantages.  If wheel saturation does ever occur, a return to zenith pointing for quick desaturation is possible.

\subsubsection{Thermal Control System}

Thermal control of the AXTAR spacecraft will utilize passive components including multilayer insulation, high
emissivity paint and coatings, heaters, etc. to maintain spacecraft subsystem components within acceptable
temperature ranges. 

\begin{figure}
\begin{center}
\includegraphics[width=5in]{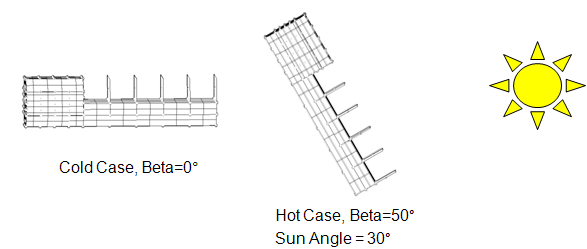}
\end{center}
\caption{Spacecraft orientations used for the hot and cold thermal cases.}\label{fig:thermal_angles}
\end{figure}

A system level thermal model was developed in Thermal Desktop\footnote{C\&R Technologies, Littleton, CO, \url{http://www.crtech.com}}. The structure is modeled as aluminum
and the panel thickness is consistent with the structural design. Structural panels double as radiative surfaces.
Environmental heat loads were calculated for an earth orbit altitude of 585 km. Spacecraft structure temperatures
for both a hot and cold orientation were generated. The hot orientation is defined as a LATA-to-sun angle of 30
degrees with a beta angle of 50 degrees; the cold orientation is defined as a LATA-to-sun angle of 90 degrees with a beta angle of 0 degrees (see Figure \ref{fig:thermal_angles}). 

Subsystems equipment and experiment heat loads are imposed directly on the structure and modeled as
area averaged heat loads. RCS thrusters, antennas, solar arrays and array mechanisms are not part of this
preliminary analysis, nor is experiment temperature prediction included. The analysis predicts that all interface
temperatures are within an acceptable range. The estimated total mass for the thermal control system is 50 kg, which includes a 30 percent margin.

\subsubsection{Propulsion System}

The propulsion system's primary function is to de-orbit the spacecraft at the end of the mission and to provide attitude control during each maneuver.  A simple monopropellant blowdown system, with maximum use of off-the-shelf components, is selected for this task.  The system consists of three ATK 80488-1 diaphragm tanks that are loaded with hydrazine propellant and nitrogen pressurant.  The thruster configuration includes four pods, each containing one Aerojet MR-104A/C (100 lbf) engine and two MR-106L (5 lbf) engines; two of the four larger thrusters are designated as backup.  This engine thrust combination is selected to maintain low gravity-losses during the de-orbit burns, allowing the use of the above-stated tanks.  All the tanks are loaded to capacity with 210.0 kg of hydrazine and 2.4 kg of nitrogen.  The basic dry mass of the entire system is 66 kg.  Adding 30\% for growth raises the dry mass to 85 kg.  The resulting total wet mass of the propulsion system is 298 kg.

\section{CONCLUSION}
\label{sec:conclusion}  % \label{} allows reference to this section

RXTE is approaching the end of its mission life, concluding a mission that has brought enormous returns in every area of X-ray timing science.  A successor mission needs to make a major advance by bringing observing capabilities beyond those of RXTE.   Only a mission with timing as its central objective will address the science goals described in Section 2.   The central technical challenge is to achieve collecting area many times larger than RXTE, extending the observed band to higher X-ray energy, and with mission architecture to identify targets (using a sky monitor), execute optimized pointing sequences, and to handle the high data rates on bright sources, all at affordable cost.  AXTAR is mission concept that will do all these things; this paper has reported the first thorough study of what is involved in producing such a mission.  It will reach the driving goals of constraining the neutron star equation of state and using plasma motions in curved space near black holes to probe black hole physics.  The same facility will support a diverse range of other investigations similar to those that made RXTE successful, but superseding RXTE by reaching lower levels of modulation, shorter timescales and higher temporal frequencies.
 
%%%%%%%%%%%%%%%%%%%%%%%%%%%%%%%%%%%%%%%%%%%%%%%%%%%%%%%%%%%%%
\acknowledgments     %>>>> equivalent to \section*{ACKNOWLEDGMENTS}       

This work was supported in part by the NRL 6.1 Base Program funding; NASA
APRA program NNG10WF45I; the MIT Kavli Instrumentation and Technology
Development Fund; and the Technology Investment Program at NASA
Marshall Space Flight Center. 

%%%%%%%%%%%%%%%%%%%%%%%%%%%%%%%%%%%%%%%%%%%%%%%%%%%%%%%%%%%%%
%%%%% References %%%%%

\bibliography{axtar}   %>>>> bibliography data in report.bib
\bibliographystyle{spiebib}   %>>>> makes bibtex use spiebib.bst

\end{document}